\documentstyle[epsf,twocolumn,prl,aps]{revtex}
\newcommand{\pnone}{{P^{\text{none}}}}

\newcommand{\pdown}{{P_\downarrow}}
\newcommand{\nup}{{n_\uparrow}}
\newcommand{\ndown}{{n_\downarrow}}
\newcommand{\erf}{{\text{erf}}}
\newcommand{\erfc}{{\text{erfc}}}
\newcommand{\hnr}{{H_{\text{nr}}}}
\def\half{{1 \over 2}}
\begin{document}

\centerline{\bf Hysteresis, Avalanches, and Noise}
\centerline{Matthew C. Kuntz, Olga Perkovi\'{c}, Karin A. Dahmen,}
\centerline{Bruce W. Roberts, and James P.
Sethna}\footnotetext{\vskip -16pt
\hskip -16pt \vbox{\parindent=0pt Matthew C. Kuntz is a Ph.D.\
candidate at the Laboratory of Atomic and Solid State Physics,
Cornell University, Ithaca, NY 14853-2501, mck10@cornell.edu; Olga
Perkovi\'{c} is with McKinsey \&
Company, olga{\_}perkovic@mckinsey.com; Karin Dahmen has just
joined the physics faculty of the University of Illinois at
Urbana-Champaign; Bruce W. Roberts is working at Starwave
Corporation in Seattle, bwr@halcyon.com; James P. Sethna is a
professor of physics at Cornell University,
sethna@lassp.cornell.edu. Details about his research
group can be found at http://www.lassp.cornell.edu/sethna/}}

\medskip

As computers increase in speed and memory, scientists are
inevitably led to simulate more complex systems over larger time
and length scales. Although a simple, straightforward algorithm is
often the most efficient for small system sizes, especially when
the time needed to implement the algorithm is included, the scaling
of time and memory with system size becomes crucial for larger
simulations.

In our studies of hysteresis and avalanches in a simple model of
magnetism (the random-field Ising model at zero temperature), we
often have found it necessary to do very large simulations.
Previous simulations were limited to relatively small systems (up to
$900^2$ and $128^3$~\cite{smallSystems}, see
however~\cite{Robbins}). In our simulations we
have found that larger systems (up to a billion spins) are crucial
to extracting accurate values of the universal critical exponents
and understanding important qualitative features of the physics.

We have developed two efficient and relatively straightforward
algorithms which allow us to simulate these large systems. The first
algorithm uses sorted lists and scales as $O(N \log N)$, and
asymptotically uses $N\times(\text{sizeof(double)+sizeof(int)})$ bytes
of memory, where $N$ is the number of spins. The second algorithm,
which does not generate the random fields, also scales in time as
$O(N\log N)$, but asymptotically needs only one bit of storage per
spin, about 96 times less than the first algorithm. Using the latter
algorithm, simulations of a billion spins can be run on a workstation
with 128MB of RAM in a few hours.

In this column we discuss algorithms for simulating the
zero-temperature random-field Ising model, which is defined by the
energy function
\begin{equation}
{\cal H} = - \! \sum_{{<}i,j{>}} J s_i s_j - \sum_i [H(t) + h_i]
s_i ,
\label{eq:Hamiltonian}
\end{equation}
where the spins $s_i = \pm 1$ sit on a $D$-dimensional hypercubic
lattice with periodic boundary conditions. The spins interact
ferromagnetically with their $z$ nearest neighbors with strength
$J$, and experience a uniform external field $H(t)$ and a random
local field $h_i$. We choose units such that $J=1$. The random field
$h_i$ is distributed according to the Gaussian distribution
$\rho(h)$ of width $R$:
\begin{equation}
\rho (h) = {1 \over {\sqrt {2\pi}} R}\ e^{-h^2 / 2R^2}.
\label{eq:rho}
\end{equation}
The external field $H(t)$ is increased arbitrarily slowly from
$-\infty$ to $\infty$. 

The dynamics of our model includes no thermal fluctuations: each
spin flips deterministically when it can gain energy by doing
so. That is, it flips when its local field
\begin{equation}
h^{\rm eff}_i = J \sum_{j} s_j + h_i + H
\label{eq:localField}
\end{equation}
changes sign. This change can occur in two ways: a spin can be
triggered when one of its neighbors flips (by participating in an
avalanche), or a spin can be triggered because of an increase in
the external field $H(t)$ (starting a new avalanche).

The zero-temperature random-field Ising model was introduced by
Robbins and Ji~\cite{Robbins} to study fluid invasion in porous
media and front propagation in disordered systems. We have used
the same model~\cite{prl1} in a different way~\cite{different} to
model noise in hysteresis loops in disordered materials. In
particular, we wish to understand Barkhausen noise in magnetic
materials with quenched disorder\cite{Bertram}. It has been found
experimentally that when an external field is gradually applied,
many materials magnetize not continuously, but in a noisy way, with
jumps (avalanches) of all sizes. (The noise can be heard by
wrapping the magnetizing material in a coil of wire and amplifying
the signal into a speaker. The signal makes a crackling noise when
a permanent magnet is brought close, quite similar to the crackling
noises heard in fires, crisped rice cereals, and crumpled paper
\cite{crackling}.) In the steepest part of the hysteresis loop,
these avalanches are found to have a power-law distribution of
sizes with an exponent $\tau\approx 1.5$. Power laws are also found
in the distribution of avalanche times with an exponent
$\alpha\approx 2$ and in the power spectrum.

The zero-temperature random-field Ising model is interesting because,
as in the disordered magnetic materials it attempts to model, the
avalanches can have a broad range of sizes. If all the avalanches were
small, understanding them would be straightforward and not very
interesting. Indeed, at large disorder $R$, the chance that a spin
which has just flipped will trigger one of its $z$ neighbors scales
roughly as $z J/R$. If this quantity is smaller than unity (large
disorder), all avalanches will be small: the noise will be a series of
small pops all of about the same size. This behavior is
uninteresting not because it is simple, but because the behavior is
strongly dependent on the details of the model at short distances,
where the model is at best a caricature of a real material.

It also is easy to understand the system in the small disorder regime
$z J/R >> 1$, where almost all the spins flip over in one infinite
avalanche. There are many problems (for example, fracture and
first-order phase transitions) where a single nucleation event leads
to the release of the stored energy in a single catastrophic event.

\begin{figure}[thb]
\begin{center}
\leavevmode
\epsfxsize=8cm
\epsffile{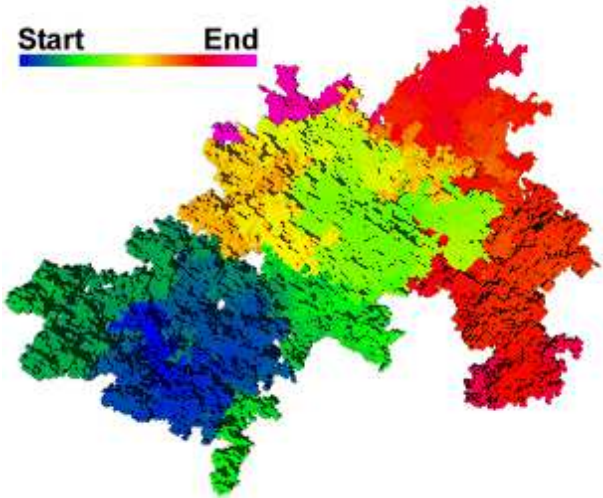}
\end{center}
\caption{A three-dimensional view from one side of a single
avalanche in a $200\times 200\times 200$ system at $R=2.3$ (within
6\% of the critical disorder $R_c$). The avalanche contains 282,785
spins. The time when each spin flipped is shown by its color. The
avalanche generally grew from left to right. Note that it has many
branches and holes; the large avalanches in three dimensions
probably have a fractal dimension a little less than three. Also
notice that on the right hand side, there are
several dark red spots poking through in the middle of the light
green area. The green area stopped growing, but other parts of the
avalanche later filled in the holes.}
\label{fig:3d_avalanche}
\end{figure}

\begin{figure}[thb]
\begin{center}
\leavevmode
\epsfxsize=8cm
\epsffile{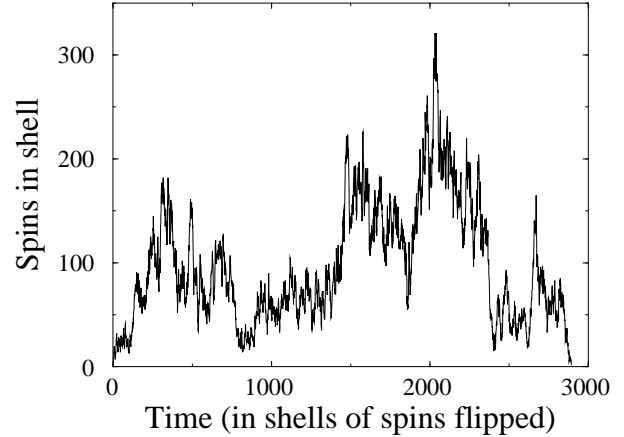}
\end{center}
\caption{A time series showing the number of spins which flipped in
each shell of the avalanche shown in
Fig.~\protect\ref{fig:3d_avalanche}~\protect\cite{LesHouches}. Note
that the avalanche is a series of bursts: near the critical point,
the avalanche is always on the verge of halting, so it proceeds in
fits and starts.}
\label{fig:time_series}
\end{figure}

\begin{figure}[thb]
\begin{center}
\leavevmode
\epsfxsize=8cm
\epsffile{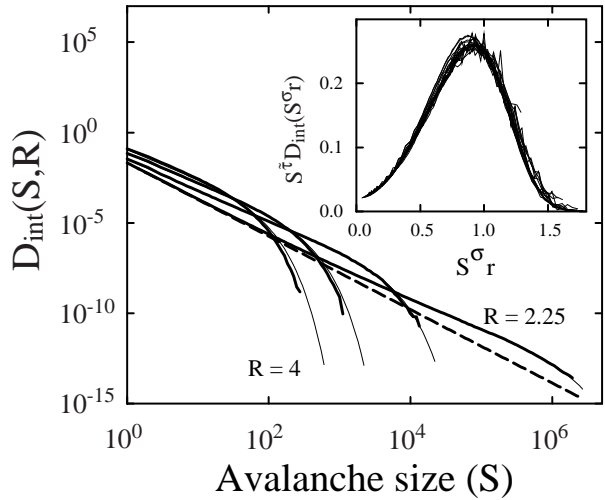}
\end{center}
\caption{Distribution of avalanche sizes for different
values of the disorder $R$ in three dimensions. Some
avalanches remain large (hundreds of spins) for $R$ a factor of two
above the critical value $R_c \sim 2.16$ where we expect a pure
power law. The avalanches are enormous (millions of
spins) when the system is still 4\% away from the critical
point; for this reason we need large systems. The inset
is a scaling collapse of the data: the thin lines in the main figure
show the scaling prediction for the avalanche sizes stemming from
the scaling collapse\protect\cite{prl3}. Note that the
scaling predictions already work well at $R=4$. The
pure asymptotic power law behavior is not yet seen at $R=2.25$,
when six decades of scaling are observed. We needed simulations of
a billion spins to show convincingly that the power law
would eventually occur\protect\cite{prl3}.}
\label{fig:avalanche_histogram}
\end{figure}

\begin{figure}
\begin{center}
\leavevmode
\epsfxsize=8cm
\epsffile{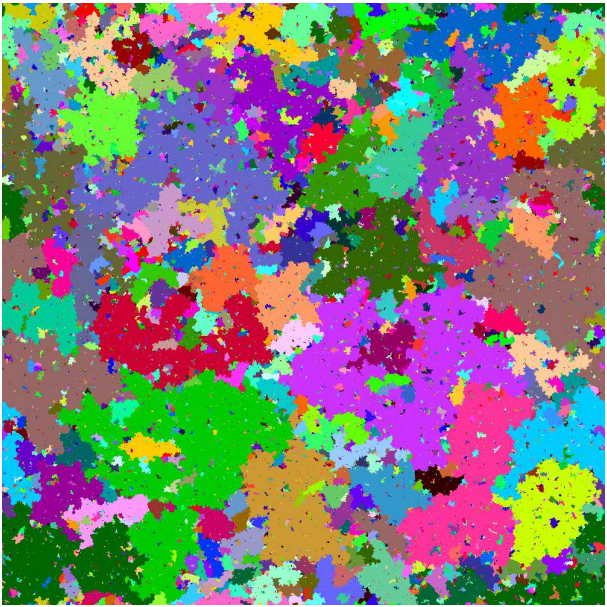}
\end{center}
\caption{A $30,000\times 30,000$
simulation with disorder $R=0.65$, where each pixel represents a
$30\times 30$ square, and each avalanche is a different color.
Note that there are avalanches of all sizes, with many smaller
avalanches, and fewer large ones.}
\label{fig:2d_avalanches}
\end{figure}

We focus on the crossover between these two limiting cases, where the
system exhibits crackling noise with avalanches of all sizes. For a
particular value of the disorder $R=R_c$, a spin which has just
flipped will on average flip exactly one neighbor as the external
field $H(t)$ is increased to a particular value $H_c$. The avalanches
at $R_c$, $H_c$ (the critical point), are finely balanced between
stopping and growing forever. They advance in fits and starts (see
Figs.~\ref{fig:3d_avalanche} and \ref{fig:time_series}) and come in
all sizes (Figs.~\ref{fig:avalanche_histogram} and
\ref{fig:2d_avalanches}) with a probability which decreases as a power
law of the number of spins in the avalanche. At $H_c$, the
distribution of avalanche sizes decays with an exponent of
$\tau\approx 1.6$ (quite close to the experimental results), and
integrated over all $H$, the distribution decays with an
exponent\cite{tildetau}
$\tilde\tau\approx 2$. Below the critical disorder
$R_c$, there will be an avalanche which will flip a nonzero fraction
of the spins in the system even as the system size goes to infinity:
we call this avalanche the infinite avalanche. There are very large
avalanches even for disorders far above the critical disorder. In
three dimensions, there are still two decades of power law scaling
50\% above the critical point. However, the convergence to the
expected asymptotic power law is very slow (Figure
\ref{fig:avalanche_histogram}). This behavior means that we see
critical scaling even if we do not fine tune $R$ to $R_c$, but we
must use very large systems to get close enough to $R_c$ to obtain a
convincing power law. In practice, we needed simulations of
approximately a billion spins to understand the physics in three
dimensions~\cite{prl3}. Two dimensions remains a challenge because
the proper scaling is not clear even for $30,000^2$
spins~\cite{prl3,Drossel}.

It is crucial with this many spins that our
algorithms be efficient both in computer time and memory. We begin
by giving the simple, but inefficient approach which has an
execution time which scales as $O(N^2)$. We then develop a more
efficient approach using a sorted list which gives an execution
time which scales as $O(N \log N)$, but which needs memory storage
which scales as
$N\times\text{(sizeof(double)+sizeof(int))}$. A billion spins would
demand 12 Gigabytes of RAM for efficient execution, which is not
usually available. Finally, we give an algorithm whose execution
time also scales as $O(N \log N)$, but whose memory requirements
are asymptotically only one bit per spin. In this case
$10^9$ spins requires 120 MB of storage, which is feasible on a
standard workstation~\cite{footnote1}. We conclude with a
discussion of time and space issues for calculating and storing
histograms and correlation functions.

\medskip \noindent {\bf The Brute Force Method}

The brute force method is the easiest one to implement and is
competitive for system sizes up to about 10,000 spins. In this
method, we store a spin direction and a random field for each site
of the lattice. We can then proceed as an experimentalist would by
measuring the magnetization at specific predetermined values
of $H$. We start with magnetization $M=-N$ and a large
negative field
$H_0$ and then increment to $H_1$, check all spins in the lattice,
and flip those spins in a positive local field. Then we must check
the neighbors of the flipped spins again to see if their local
fields are now positive. This procedure is continued until all the
neighbors of flipped spins have been checked. We then repeat the
whole procedure again for a new field
$H_2$, and so on. This approach gives the correct magnetization at
the fields
$H_n$: the order in which spins are flipped can be shown not to
influence the final state~\cite{Middleton,prl1}. However, unless the
increments in $H$ are very small, several avalanches
may occur in a given increment, and all information about single
avalanches (such as histograms of avalanche sizes) will be
distorted.

The time for the brute force method scales as $O(N X T$), where
$X$ is the number of fields $H_n$ at which the magnetization is
measured, and $T$ is the average time needed to check the neighbors
of the flipped spins measured in units of shells of neighbors.

\begin{figure}[thb]
\begin{center}
\leavevmode
\epsfxsize=8cm
\epsffile{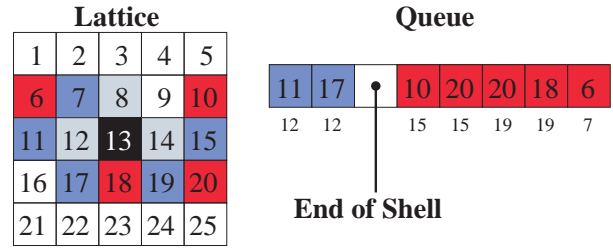}
\end{center}
\caption{Example of how a queue is used to
propagate an avalanche. The colored spins are spins which either
have flipped in the current avalanche or will flip in the current
avalanche. Spin 13 triggered the avalanche, then the light grey
spins (14,8,12) were put on the queue as the first shell. As they
flipped, the second shell, the blue spins (15,19,7,11,17), were
put on the queue. As the first blue spins (15,19,7) flipped, the
dark red spins (10,20,20,18,6) were added to the queue as the
start of the third shell. The next spin to flip is at the left
hand side of the queue. When this spin flips, its neighbors will
be checked, and the neighbors which are ready to flip will be
added to the right hand side of the queue. The small numbers
below the spins in the queue indicate which neighbor caused the
spin to be put on the queue. Note that different neighbors can
cause a spin (such as spin 20) to be put on the queue more than
once. We have to be careful to only flip the spin once.}
\label{fig:avalanche_propagation}
\end{figure}

A variation on this approach is to propagate one avalanche at a time
as shown in Fig.~\ref{fig:avalanche_propagation}:
\begin{enumerate}
\item Find the triggering spin for the next avalanche by checking
through the lattice for the unflipped site with the largest internal field
$h_i^{\rm int} = h^{\rm eff}_i - H$. 
\item Increment the external field so it is just large enough to flip the site,
and push the spin onto a first-in first-out (FIFO) queue
(see Fig.~\ref{fig:avalanche_propagation}, right).
\item Pop the top spin off the queue.
\item If the spin has not been flipped, flip it and push all
unflipped neighbors with positive local fields onto the queue.
\item While there are spins on the queue, repeat from step 3.
\item Repeat from step 1 until all spins are flipped.
\end{enumerate}

This method is standard for avalanche propagation problems. It
also is related to the propagation of cluster flips in the Wolff
algorithm~\cite{Wolff}. Using a queue instead of recursion has two
advantages. First, recursion is slower and more memory
intensive, because each recursive call must push all local
variables and all registers onto the system stack (which usually
has a pre-allocated limit). If we use our own queue, we need only
to push the coordinate of the next spin on the queue each time, and
we can make the queue as large as necessary. Second, in order to
produce a natural spin-flip order, we want to flip all spins that
are ready to flip at a given time {\em before} we flip the spins
that they cause to flip. If we put spins that are ready to flip on
a FIFO queue, we correctly achieve this order. This
procedure corresponds to doing a breadth-first search. Recursion,
which is the same as putting the spins on a LIFO stack, would
explore all possible consequences of flipping the first neighbor it
looks at before it considers the second neighbor. This
depth-first search produces an unnatural spin-flip order (although
the final state after the avalanche is
unchanged~\cite{Middleton,prl1}). The dynamics during the avalanche
of Fig.~\ref{fig:time_series} assumes one shell of spins flipped
during each time slice, which is easy to determine by placing
markers on a FIFO spin queue, as shown in
Fig.~\ref{fig:avalanche_propagation}. Each time the marker
is popped off of the queue, a new shell is started and the
marker is put back on the end of the queue.

Doing the brute force algorithm one avalanche at a time is
very inefficient except at very low disorders. Sweeping
through the entire lattice for each avalanche takes $O(N)$ time per
avalanche. Because there are $O(N)$ avalanches, the total running
time scales as $O(N^2)$. A hybrid approach, finite steps in field
followed by internal propagation of avalanches, could be quite
efficient if one is solely interested in the magnetization at those
fields. A brute force method is probably necessary when simulating
systems with long-range interactions~\cite{Zapperi}.

\vfil\break \noindent {\bf Time Efficiency: Sorted Lists}

The brute force method is very inefficient at locating the origin of
the next avalanche, and we are immediately led to think of storing
the several largest local fields in each sweep. If we take this
thinking to its logical conclusion, we are led to store a list of
all of the spins in the system, sorted according to their random
fields.

\begin{figure}[thb]
\begin{center} 
\leavevmode
\epsfxsize=8cm
\epsffile{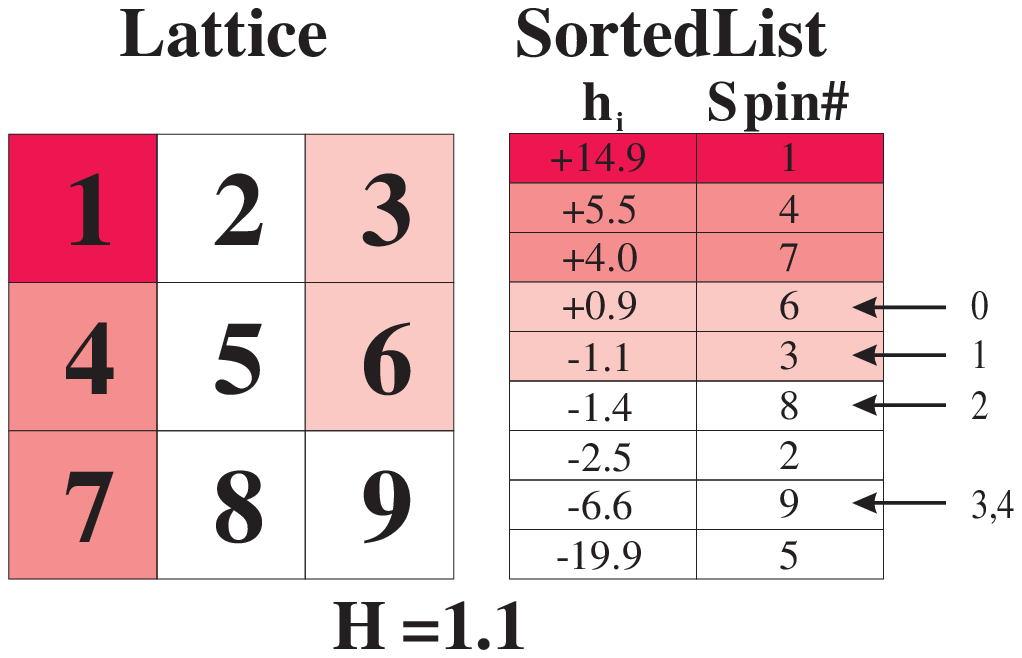}
\end{center}
\caption{Example of how a sorted list is used to find the
next spin in the avalanche. The colors indicate spins which have
already flipped. The first column in the sorted list contains the
random field, and the second column contains the number of the spin
with that random field. The arrows to the right indicate the
nextPossible[$\nup$] pointers --- the first spin which would not
flip with $\nup$ neighbors up. The spins pointed to are the
possible starting locations for the next avalanche. Note that
some of the pointers point to spins that have already flipped,
meaning that these spins have more than $\nup$ neighbors
up. In a larger system, the unflipped spins will not all be
contiguous in the list.}
\label{fig:sorted_list}
\end{figure}

Unfortunately, life is complicated by the fact that spins experience
not only their local random fields, but also fields from their
neighbors. To find the origin of the next avalanche, we use the
following algorithm:

\begin{enumerate}
\item Define an array $\text{nextPossible}[\nup]$, $\nup = 0,1\ldots
z$, which points to the location in the sorted list of the next spin
which would flip if it had $\nup$ neighbors. Initially, all the
elements of $\text{nextPossible}[\nup]$ point to the spin with the
largest local random field, $h_i$.
\item Choose
from the $z+1$ spins pointed to by $\text{nextPossible}$, the one
with the largest internal field $h_i^{\rm int}[\nup] = \nup -
\ndown + h_i = 2\nup - z + h_i$.
\item Move the pointer $\text{nextPossible}[\nup]$ to
the next spin on the sorted list. 
\item If the spin with the largest $h_i^{\rm
int}[\nup]$ has
$\nup$ up neighbors, then flip it. Otherwise go back to step 2. 
\end{enumerate}
\noindent An example of the sorted list and the pointers from
$\text{nextPossible}$ is shown in Fig.~\ref{fig:sorted_list}.

The sorting of spins can be done in time $O(N \log N)$. Storage with
this algorithm is $N\times\text{sizeof(int)}$ for the sorted array (if
we reduce the $D$-dimensional coordinates to one
number~\cite{footnote3}), and $N\times\text{(sizeof(spin) +
sizeof(double))}$ for the lattice itself. Various other compromises
between speed of execution and storage are possible, but all leave the
running time $O(N\log N)$. The sorted-list algorithm is fast: the
largest system sizes we can store on a reasonable workstation execute
$1000^2$ and $100^3$ spins in a few seconds. It is the method of
choice for these small systems or when one is interested in the
behavior for non-monotonically increasing
fields.~\footnote{The sorted-list algorithm can be used for
non-mono\-ton\-ically increasing
fields with only a few minor additions. When the external field is
being lowered instead of raised, the avalanche propagation is the
same, except spins are flipped when their local field becomes less
than zero instead of when it becomes greater. The nextPossible array
needs to be handled carefully. The next spin that would flip up if the
field were increased is the last spin that would have already flipped
down with the field decreasing. Every time the direction of change of
the external field is reversed, all of the nextPossible[$\nup$]
pointers need to be adjusted by one to account for this.}

\medskip \noindent {\bf Space Efficiency: One Bit per Spin}

The combination of the rapid execution of the sorted-list algorithm
and large finite size effects led us to develop an algorithm optimized
for memory efficiency. The key is to recognize that we need never
generate the random fields! In invasion percolation~\cite{invasion}
(and in the interface problem~\cite{Robbins} analogous to ours) 
the random fields are generated only on sites along the boundary of
the growing cluster. In our problem, we can take this idea further:
for each change in a spin's local field given by
Equation~\ref{eq:localField}, we generate only the probability that
it will flip. Storing the random fields is unnecessary because the
external field, the configuration of the spin's neighbors, and the
knowledge that the spin has not yet flipped gives us all the
information which we need to determine the probability that the spin
will flip. The only quantity which we must store for each site of
the lattice is whether the spin is up or down. Thus, we can store
each site of the lattice as a computer bit saving large amounts of
memory. 

\begin{figure}[thb]
\begin{center}
\leavevmode
\epsfxsize=8cm
\epsffile{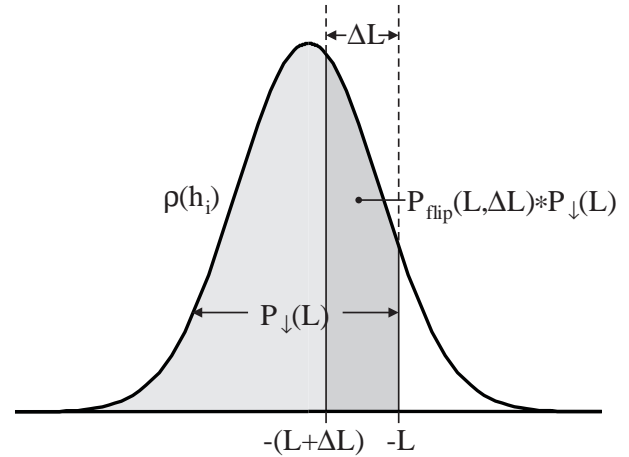}
\end{center}
\caption{The probability that a spin will not have flipped by the
time its local field reaches $L$ is the probability that the random
field is less than $-L$. This probability is represented by the
shaded area of the Gaussian. The probability that the spin will
flip before the field reaches $-(L+\Delta L)$ is represented by the
area of the darker region divided by the area of the shaded
region.}
\label{fig:gaussian}
\end{figure}

For a monotonically increasing external field, the conditional
probability that a spin flips before its nonrandom local field,
$\hnr\equiv H+(2\nup-z)$, reaches $\hnr+\Delta \hnr$ given that it has
not flipped by $\hnr$ is 
\begin{equation}
\label{eq:ProbFlipped}
P_{\text{flip}}(\hnr,\Delta \hnr) =
\frac{\left[\pdown(\hnr)-\pdown(\hnr+\Delta
\hnr)\right]}{\pdown(\hnr)},
\end{equation}
where $\pdown(\hnr)$ is the probability that a spin points
down when the local field is $\hnr$. A spin with local
field
$\hnr$ will point down if its random field $h_i$ satisfies
$h_i+\hnr \leq 0$. This condition implies that the probability
that a spin with $\nup$ up neighbors points down is
\begin{eqnarray}
\label{eq:ProbNotFlipped}
\pdown(\nup,H) &=& 
	\! \int_{-\infty}^{-\hnr(\nup,H)} \rho(h)\, dh \\
		&=& \half + \half \erf\left(-\hnr(\nup,H)/\sqrt{2} R\right)
			\nonumber\\
\label{eq:ProbNotFlipped2}
		&=& \half \erfc\left(\hnr(\nup,H)/\sqrt{2} R\right).
\end{eqnarray}
(Writing $\pdown$ in terms of the $\erfc$ function removed some
problems with rounding at large negative fields $H$.) These
probabilities are illustrated graphically in
Fig.~\ref{fig:gaussian}.

Finding the next avalanche is subtle when the random fields are not
stored: changing the external field $H$ introduces a probability that
any unflipped spin in the lattice may flip. Inspired by the
continuous time Monte-Carlo algorithm~\cite{BKL}, we keep track of
$N_{\nup}$, the number of down spins which have $\nup$ up
neighbors. Given the probabilities that spins with $\nup$ up
neighbors will flip, we calculate both the change in the external
field
$\Delta H$ needed to flip the next spin and the probability that
the next spin to flip has $\nup$ up neighbors. We then randomly
choose $\nup$ and search at random through the lattice for a spin
with $\nup$ up neighbors. The time taken for the search is the part
of the algorithm which scales worst for large $N$. If there are
$N_{\nup}$ spins left, this search will take an average time
$O(N/N_{\nup})$. Summing over
$N_{\nup}$ and $\nup$ yields a bound of order $z N \log N$. In one
of our programs, we use a tree structure to do this search more
efficiently; this complication decreases the running time by 40\%
for a $500^2$ system at $R=1$.

How do we calculate $\Delta H$? From Equation~\ref{eq:ProbFlipped},
the probability that a single spin with $\nup$ neighbors up has not
flipped in the range $\Delta H$ is $1-P_{\text{flip}}(\nup,H,\Delta H) =
\pdown(\nup,H+\Delta H)/\pdown(\nup,H)$. The probability that
no spin with $\nup$ up neighbors has flipped in this range is
\begin{equation}
P_{\nup}^{\rm none} = 
	 \left[ \frac{\pdown(\nup,H+\Delta H)}{\pdown(\nup,H)}
\right]^{N_{\nup}},
\label{eq:ProbNoneFlippedNUp}
\end{equation}
and the probability that no spin has flipped between $H$ and $H+\Delta
H$ is
\begin{equation}
P^{\rm none}(\Delta H) = \prod_{\nup=0}^z P_{\nup}^{\rm none}.
\label{eq:ProbNoneFlipped}
\end{equation}
To find $\Delta H$, we choose a random number $r$ uniformly
distributed between zero and one, and set $\Delta H$ so that
$\pnone(\Delta H) = r$.

Unfortunately, we cannot solve for $\Delta H$ analytically, and we
must find a numerical solution. To find this
solution efficiently, we need a good initial guess. In
analogy with nuclear decay, if spins flip with a constant rate
$\Gamma$, we expect the probability that no spins have yet flipped
to be
$e^{-\Gamma
\Delta H}$. So, for a first approximation of $\Delta H$, we assume
that the spin-flip rate $\Gamma$ is a constant, and therefore
\begin{equation}
\Delta H_1 = -\frac{\log({r})}{\Gamma(H)}.
\label{eq:DeltaHGuess}
\end{equation}
where $\Gamma(H)$ is given by
\begin{eqnarray}
\Gamma(H) &=& -\frac{d \log(\pnone(\Delta H=0))}{d \Delta H} \nonumber
\\
	&=& -\sum_{\nup=0}^z N_\nup \frac{d
\log(\pdown(\nup,H+\Delta H))}{d\Delta H} \nonumber \\
	&=& \sum_{\nup=0}^{z} N_\nup \frac{\rho
(\hnr(\nup,H))}{\pdown(\nup,H)} \equiv \sum_{\nup=0}^{z}
\Gamma(\nup,H)
\label{eq:Gamma}
\end{eqnarray}
We can make a better second guess by looking at the error in our first
guess. If the error in our guess is $\Delta r = P^{\rm
none}(\Delta H) - r$, then we can make an improved second
guess for $\Delta H$ by aiming for $r-\Delta r$:
\begin{equation}
\label{eq:DeltaHGuess2}
\Delta H_2 = -\frac{\log({r-\Delta r})}{\Gamma(H)}.
\end{equation}
These two guesses can then be used as input into a root finding
routine~\cite{footnote4}. Note that while these guesses are usually
very good for small $|H|$ and lead to quick solutions, they can be
very bad for large $|H|$. If the guesses for $\Delta H$ are very
large, it may be better to choose two arbitrary guesses. In our
code, if $\Delta H_1>20$, we use $\Delta H_1=0$ and $\Delta H_2=20$
for the two guesses.

Our algorithm for finding the next avalanche becomes
\begin{enumerate}
\item Choose a random number $r$ uniformly distributed between
zero and one.
\item Pre-calculate the values of $\pdown(\nup,H)$ using 
Equation~\ref{eq:ProbNotFlipped2}. These values will be used repeatedly
in solving for $\Delta H$.
\item Calculate guesses for $\Delta H$ using
Equations~\ref{eq:DeltaHGuess} and \ref{eq:DeltaHGuess2}, and use them
as input to a root finding routine to find the exact solution for
$\Delta H$.
\item Increment $H$ by $\Delta H$.
\item Calculate the array probFlip[$\nup$] for use in the remainder
of the avalanche, where probFlip[$\nup$] is the probability at the
current field $H$ that a spin will flip when its number of up
neighbors changes from $\nup$ to $\nup$+1 (see
Equation~\ref{eq:ProbFlipped}).
\item Calculate the rates for flipping spins for each $\nup$ {\em at the
current field}
$H=H_{\rm old}+\Delta H$:
\begin{equation}
\Gamma(\nup,H) = N_\nup \rho(\hnr(\nup,H))/\pdown(\nup,H)
\label{eq:gammas}
\end{equation}
and the total rate $\Gamma(H) = \sum_{\nup=0}^{z} \Gamma(\nup,H)$.
\item Choose a random number uniformly distributed between
zero and $\Gamma$ and use it to select $\nup$.

\item Search at random in the lattice for an unflipped spin with
$\nup$ up neighbors~\cite{footnote5}.
\item Start the avalanche at that spin. During an avalanche, the
algorithm is essentially the same as the brute force algorithm:

\item Push the first spin onto the queue.
\item Pop the top spin off of the queue.
\item If the spin is unflipped, flip it, find $\nup$, and decrease
$N_\nup$ by one. Otherwise, skip to step 14.
\item Look at all unflipped neighbors. For each unflipped neighbor,
find the current number of up neighbors, $\nup$; decrease $N_{\nup-1}$
by one, and increase $N_{\nup}$ by one. Push the spin on the
queue~\cite{footnote6} with probability probFlip[$\nup-1$], as
calculated in step (5).
\item While there are spins left in the queue, repeat from step 11.
\item While there are unflipped spins, repeat from step 1.
\end{enumerate}

This algorithm is about half as fast in practice as the sorted-list
algorithm which is faster than we expected. The overhead involved
in solving for $\Delta H$ is presumably compensated by the time
saved not shifting data in and out of cache. Systems of
$10^9$ spins take a few days of CPU time on a reasonable
workstation;
$30,000^2$ systems take less than 15 hours on a 266 MHz Pentium II.

\medskip \noindent {\bf Calculating Histograms and Correlations}

Several functions are needed to characterize the critical
properties of our model. The simplest function is the
magnetization
$M$ as a function of the external field
$H$. We also calculate distributions of avalanche sizes
(Fig.~\ref{fig:avalanche_histogram}, top), and correlation
functions. Some care must be taken to make sure that the
calculation of these functions does not dramatically increase the
running time or memory requirements of the simulation.

When doing calculations with a billion spins, we cannot output
any quantity which scales linearly with the system size.
Instead of computing $H(M)$ at each avalanche (about a GB of 
data, which rapidly would fill our disk), we are forced to compute
$H(M_n)$ at pre-chosen points.

The characteristic feature of the critical point is the appearance
of an infinite avalanche. The equivalent of an infinite avalanche
in a finite system is an avalanche which spans the entire system in
at least one dimension. To tell whether we are above or
below the critical point, we need to detect these spanning
avalanches. In three and higher dimensions, the number of
spanning avalanches as a function of $R$ is also
interesting to study. The most obvious way of detecting spanning
avalanches is to mark each row as a spin flips in it, and check at
the end of the avalanche to see if all the rows contain flipped spins
from the avalanche. However, this method requires O($N^{1/D}$)
operations per avalanche. Because there are many small avalanches,
this method is unacceptable. A preferable method is to keep track of
the
$2 \times D$ boundaries of the avalanche as it grows. If a pair of
boundaries meet, then the avalanche is a spanning avalanche. We
must take care to treat the periodic boundary
conditions properly.

Another useful function is the avalanche size
distribution $D(S)$, defined as the number of avalanches which flip
$S$ spins during the simulation, divided by the total number of
spins. Like the $M(H)$ curve, the avalanche size
distribution scales linearly with the system size. Thus, we
need bins up to size $N$, the size of the largest possible
avalanche. Logarithmic binning is the obvious solution, with bin
$n$ including all sizes $b_a^{n-1} < S < b_a^n$. We have chosen
$b_a$ from 1.01 to 1.1. Large bins are preferable for lower
statistical noise. This choice is particularly important in the tail
of very large avalanches, where small bins would contain few
avalanches. However, very large bins will systematically alter
the shape of the scaling functions (although they will not change
the critical exponents). It is important to divide the final
population in each bin by the number of integers contained within
the bin (and not just the bin width). Clearly we should also
ignore the early bins which do not contain any integers.

We calculate the correlation function $G(x,R)$
within an avalanche, where $G(x,R)$ gives the probability that
the first spin in an avalanche will cause a spin a distance
$x$ away to flip in the same avalanche. At the beginning of each
avalanche, we record the coordinates of the first spin in the
avalanche. Then, for each subsequent spin in the avalanche, we
calculate the distance $x$ to the first spin, and add one to the
appropriate bin. Logarithmic binning is not necessary for the
correlation function, because the size of the correlation function
is proportional to the length of the system, not the total number
of spins. Thus, we use a fixed bin size $b_c=1$. At the end of the
simulation, each bin should be normalized by the number of spins
which are between
$x-b_c/2$ and
$x+b_c/2$ away from the origin.

The only tricky part of calculating $G(x,R)$ comes from the
periodic boundary conditions. If the avalanche crosses a boundary,
two points at opposite ends of the avalanche can come close
together. Because we do not calculate $G(x,R)$ for spanning
avalanches, we know that there will be at least one row in every
dimension which is not touched by the avalanche. To calculate
separations, we use the periodicity of the lattice and the
continuity of the avalanche to shift the coordinates so they are
all on one side of these empty rows. Because we are already keeping
track of the boundaries of the avalanche for the detection of
spanning avalanches, finding an empty row is easy.

\begin{figure}[thb]
\begin{center}
\leavevmode
\epsfxsize=8cm
\epsffile{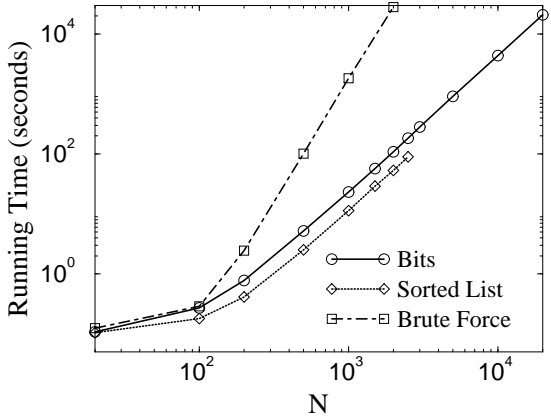}
\end{center}
\caption{The running times for the three algorithms for
two-dimensional systems with $R=1.0$ on a 266 MHz Pentium
II with 128 MB of memory. Note that both the bits algorithm
and the sorted-list algorithm have run times which grow
approximately linearly (the $\log N$ is not visible), and the
brute force running time grows quadratically. Also notice that
the largest bits simulation was 64 times larger than the
largest sorted-list simulation.}
\label{fig:running_time}
\end{figure}

The running times of the three algorithms as a function of system
size are shown in Fig.~\ref{fig:running_time}. The brute force
algorithm can be useful when one cares only about $M(H)$ at a few
points, but is otherwise too slow for large systems. The
sorted-list algorithm is the fastest algorithm, but on a 128 MB
machine, only system sizes of about six million spins can be run. 
The bits algorithm is almost as fast as the sorted-list algorithm,
and asymptotically uses only one bit of memory per spin.

We have made C++ source code implementing all three algorithms
available at\hfil\break
http://www.lassp.cornell.edu/sethna/hysteresis/code/.
Each algorithm is contained in a separate, well-documented class.
There are classes for detecting spanning avalanches, measuring
avalanche size distributions, and measuring correlation functions. 
There are also both Microsoft Windows and command line interfaces
to the code. The command line interface should be portable to any
computer with a C++ compiler and an implementation of the Standard
Template Library. We have also compiled executables for
Windows 95/NT, Linux, and several other flavors of UNIX. The
running times in Fig.~\ref{fig:running_time} all come from the
code compiled under Linux.

Working on hysteresis, avalanches and noise with our model has been
very rewarding. The simulations are beautiful and entertaining in
themselves. Developing faster and more space efficient algorithms
was amazingly satisfying: each new method not only eased our lives
and our computer budgets, but also opened a whole new window on the
behavior of the model. It was also fun developing new ways of
measuring what was happening in the simulations: watching the spins
flip and measuring the avalanche size distribution was only the
beginning. We have written a few exercises which we hope will
entertain and inform you as they did us.

\medskip \noindent {\bf Suggested problems for further study}

\smallskip \noindent 1. {\it Phase Transitions in the Shape of the
Loop}. Download our program from\hfil\break
http://www.lassp.cornell.edu/sethna/hysteresis/code/. Also download
DynamicLattice and xmgr if you are using Unix. Run the program.
Under Windows, just press OK in the opening dialog box to run a
simulation with the default parameters. Under UNIX, type {\tt run}
at the {\tt >>>} prompt. (More detailed instructions can be found
on our Web site.) Try the other two algorithms. For smaller
systems, brute force works acceptably, but for $L=500$ it is rather
slow.

You should see an animation of the avalanches as the external field
is ramped upward and the spins flip. After the simulation ends, you
should obtain a graph of $M(H)$, the avalanche size distribution
$D(S,R)$, and the correlation function $G(x,R)$.
The $M(H)$ curve shows the bottom half of the hysteresis loop: it
should consist of many jumps of various sizes. The top half of the
hysteresis loop is pretty much the same shape $-M(-H)$, but the
details of the jumps are different.

The avalanche size distribution $D(S)$ measures the number of jumps
as a function of size $S$. It should look like a fairly good power
law, and be a straight line on a log-log plot.

A log-log plot of the correlation function $G(x,r)$ looks much less
linear. At small distances, the correlation function decreases
as a power law, but the power law behavior bends over after only a
decade or so. This behavior is a symptom of $R$ not being quite
at the critical disorder
$R_c$.

(a) According to our scaling theory~\cite{prl1,prl3}, near the
critical point
$D(S,R_c) \sim S^{-\tilde\tau}$. What is your best estimate for
the exponent $\tilde\tau$ at the default value $R=1.0$? 

(b) Do a simulation at a smaller value of disorder $R$, say
0.8. Does the behavior of $G(x,R)$ remain a power law over larger
values of
$x$?

(c) Do a simulation with $D=3$, $L=50$, and $R=2.5$. What is
$\tilde\tau$ in three dimensions? (To obtain better data,
do several runs and use the averaging option.) Consider $R=2.1$.
Does the shape of the $M(H)$ curve look qualitatively different? We
believe that the power law distributions occur for $R_c \sim 2.16$.
At this value, the hysteresis loop first develops a macroscopic jump
(the infinite avalanche). One obtains power laws and scaling at the
phase transition in the shape of the hysteresis loop, between
smooth loops and ones with an infinite avalanche.

\smallskip \noindent 2. {\it Time, Space, and Bits}. We have a
local, somewhat older supercomputer with four gigabytes of RAM. How
large a system could we run in this memory using the three
algorithms we discussed? Ignore all the memory requirements except
those that scale linearly with the number of spins
$N$; the rest is negligible. Assuming the time spent in bits
continues to grow as
$N \log N$ after the last data point in
Fig.~\ref{fig:running_time}, how long will the largest possible
bits simulation take to complete?

\smallskip \noindent 3. {\it Programming}.
To do the following problems, you will need to download the source
code for our program from\hfil\break
http://www.lassp.cornell.edu/sethna/hysteresis/code/. You will also
need a C++ compiler which supports the standard template libraries.
The standard template libraries are an important advance for
scientific programming. We find them incredibly useful, and look
forward to their wide implementation now that the C++ standard is
in place. Links to compilers supporting the standard template
libraries can be found on our Web page. 

Run the simulation described in Problem 1 to test that the
program works. Then try the following problems (many details on how
to work with our source code can be found on our Web site).

(a) Adding a Spin-Flip Action: Time Series for Large
Avalanches. We
have designed our code so that it is easy to add new types of
measurements. One quantity which experimentalists measure is the
change in magnetization with time. In our code, we record the time
series of the whole run, with each avalanche represented as a
single point. There is also interesting structure within each
avalanche (see Fig.~\ref{fig:time_series}). Add a new class to the
program which records the time series {\em within} the largest
avalanche.

(b) Implementing Brute Force. Implement the
brute force algorithm. You can either replace the {\tt
BruteForceHysteresisSimulation} class or implement it from scratch.

(c) Implementing Sorted Lists. Implement the sorted list
algorithm by replacing the {\tt SortedListHysteresisSimulation}
class.

(d) Implement the bits algorithm.

More information on how to write your own {\tt
BruteForceHysteresisSimulation}, {\tt
SortedListHysteresisSimulation}, and {\tt BitsHysteresisSimulation}
classes are available on our Web site.

\medskip \noindent {\bf Acknowledgments}. Portions of this work
were supported under NSF DMR-9805422 and DOE DEFG02-88-ER45364. An
equipment grant from Intel and the support of the Cornell Theory
Center are also gratefully acknowledged. We thank Harvey Gould and
Jan Tobochnik for careful and helpful editing suggestions.


\begin{references}
\bibitem{different} In our model, the spins can flip anywhere in the region.
In the fluid invasion problem, the spins can only flip next to a flipped
region: the water can push into the rock only along an invading front. Our
model allows the interaction of many different boundaries between up and down
spin regions.
\bibitem{footnote1} Neither of the useful algorithms parallelizes
well; our group has made effective and extensive use of parallel
machines to average over disorder. The fluctuations at the critical
point are severe at all system sizes.
\bibitem{tildetau} The exponent $\tilde \tau$ can be written in terms of
other critical exponents for our problem: 
$\tilde \tau = \tau + \sigma \beta \delta.$
\bibitem{footnote2} Search only those spins for which
nextPossible[$\nup$] has not run off the end of the list.
\bibitem{footnote3} $(X_1,X_2,\dots,X_D)$ can be reduced to $X_1+X_2
L+\dots+X_D L^{D-1}$. This reduction yields the numbering scheme
shown in Figs.~\protect\ref{fig:avalanche_propagation} and
\protect\ref{fig:sorted_list} and also allows us to
store a $D$-dimensional lattice in a one-dimensional array and
write code which will work in any dimension.
\bibitem{footnote4} We use a simple algorithm to make sure that the
guesses are on opposite sides of the correct guess, and then use the
Brent algorithm to find the root. The Brent algorithm uses a
combination of interpolation when the guesses are close, and
bisection when the interpolation works badly. Descriptions and
implementations of this and other root finding algorithms can be
found in Refs.~\protect\cite{NRec} and \protect\cite{NetLib}.
\bibitem{footnote5} If your random number generator is not good
enough, or if you have incorrectly kept track of the number of down
spins with $\nup$ neighbors up, your simulation may enter an
infinite loop searching for the next spin. Also, if you are storing
your spins in a 1D array, you should still generate $D$ random
numbers, one for each coordinate. If you try to save time and only
generate one random number, then for a very large system there will
be gaps in the random numbers, and some spins will never be found,
causing an infinite loop.
\bibitem{footnote6} One might worry that the queue will grow very
large. However, the queue will only be as large as the largest
shell of spins, which should scale as the surface area of the
largest avalanche. Therefore, it should scale as $N^{1/2}$ in two
dimensions, and as $N^{2/3}$ in three dimensions. Because the
avalanches tend to grow in fits and starts, this situation is even
better. Note that in Fig.~\protect\ref{fig:3d_avalanche}, even
though the avalanche contained almost 300,000 spins, and the
surface area was very large, the largest shell had only 321 spins
in it (Fig.~\protect\ref{fig:time_series}). The queue is generally
very small compared with the system size.
\end{references}
\end{document}